\documentclass[aps,prd,showpacs,preprintnumbers,nofootinbib,twocolumn]{revtex4}

\usepackage{bm}
\usepackage{amssymb}

\newcommand{\be}{\begin{equation}}
\newcommand{\e}{\end{equation}}
\newcommand{\bear}{\begin{eqnarray}}
\newcommand{\ear}{\end{eqnarray}}
\newcommand{\nline}{\nonumber \\}
\newcommand{\f}{\frac}

\newcommand{\sech}{{\rm sech}}

\begin{document}

\title{Quasi normal modes in Schwarzschild-DeSitter spacetime: 
A simple derivation of the level spacing of the frequencies}

\author{T. Roy Choudhury}
\affiliation{SISSA/ISAS, via Beirut 2-4, 34014 Trieste, Italy}
\email[]{chou@sissa.it}
\homepage[]{http://www.sissa.it/~chou}
\author{T. Padmanabhan}
\affiliation{IUCAA, Ganeshkhind, Pune 411 007, India}

\date{\today}

\begin{abstract}
It is known that  the imaginary parts of the 
quasi normal mode (QNM) frequencies for the
Schwarzschild black hole are evenly spaced with a spacing that
depends only on the surface gravity. On the other hand, 
for massless minimally coupled scalar fields, there exist no QNMs 
in the pure DeSitter spacetime.  
It is not clear what the structure of the QNMs would be 
for the Schwarzschild-DeSitter (SDS) spacetime, which is characterized 
by two different surface gravities.
We provide a simple derivation of the imaginary parts of the QNM frequencies
for the SDS spacetime
by calculating the scattering amplitude 
in the 
first Born approximation and  determining its poles.
We find that, for the usual 
set of boundary conditions in which the incident wave is scattered 
off the black hole horizon, the imaginary parts of the 
QNM frequencies have a equally spaced structure  
with the level spacing depending on the surface gravity of 
the black hole. 
Several conceptual issues related to the QNM are 
discussed in the light of this result and comparison 
with previous work is presented.

\end{abstract}
\maketitle

\section{Introduction}

It is well known that the quasi normal modes (QNMs, hereafter) are 
crucial in studying
the gravitational and electromagnetic 
perturbations around black hole spacetimes 
(for comprehensive reviews and exhaustive list of references 
see \cite{nollert99,ks99}). 
The QNMs also seem to have an observational significance as 
the gravitational waves produced by the perturbations can, 
in principle, be used for 
unambiguous detection of black holes. 
The early studies 
of the QNMs concentrated on numerical computations  
(see, for example, \cite{vishveshwara70,press71}) as 
analytical solutions of the perturbation equations were difficult to obtain. 
Essentially, one solved the perturbation equations 
numerically for different initial conditions in a 
given black hole spacetime. It was found that the results were, 
in general,  
independent of the initial conditions and depended mostly on the 
parameters characterizing the black hole horizon.
For example, for the Schwarzschild black hole, the spectrum of the
QNMs was found to have the structure 
\cite{nollert93,andersson93,hod98,ga03,cly03,xsws03}
\be
k_n =   i\kappa\left(n+\frac{1}{2}\right)+\frac{\ln 3}{2\pi} \, 
\kappa  + \mathcal{O}[n^{-1/2}] 
\label{eq:kn_num}
\e
which depends only on the surface gravity of the black hole $\kappa$. 
Similarly, for charged (Reissner-Nordstr\"{o}m) 
and rotating (Kerr) black holes, 
the spectrum was found to depend only on the 
parameters characterizing the horizon, i.e., 
the mass, charge and angular momentum of the black hole 
\cite{detweiler80,mashhoon85,leaver85,leaver90,kokkotas91,aas94,omoi96,klpa97,onozawa97,bcko03,bk03,hod03,hod03b}.

The numerical studies were followed by a series of 
analytical calculations which were mainly based upon some 
approximation scheme, like for example, 
(i) approximating the scattering potential 
with some simple functions so that the problem becomes exactly solvable
\cite{fm84,bm84,molina03}, 
(ii) using WKB-like techniques 
\cite{sw85,iyer87,iw87,ks88,gwks90,si90,gm92,sw92,leaver92b,al92,ah03,konoplya03}, 
(iii) Born approximation 
\cite{mmv03,padmanabhan03b} 
and (iv) others 
\cite{leaver86,pm87,mp89,ffah92,ns92,aas93,aas93b,ans93,lm96,clsy95,clsy95b,pz99,solodukhin03,birmingham03}. 
Nearly all of these schemes came with 
their own limitations and rarely reproduced the exact form of the 
QNM spectrum as given in equation (\ref{eq:kn_num}) 
-- in particular, there was hardly 
any explanation for the $\ln 3$ term. 
Analytical proofs of equation (\ref{eq:kn_num}) 
were provided only recently, using 
the methods of continued fractions 
\cite{leaver85,leaver92,nollert93,motl02} and 
also by studying the monodromy of the perturbation 
continued to the complex plane \cite{mn03,neitzke03}.

While most of the analytical work concentrated on the real part of 
QNM (due to the interest in the
$\ln 3$ factor),  the structure of the imaginary part is equally 
intriguing and requires physical understanding.
In particular, we need to understand  why  the
imaginary part of $k_n$ in equation (\ref{eq:kn_num}) has a simple, 
equally spaced 
structure, with $\kappa$
determining the level spacing. 
Physical quantities with a quantized spectrum are always of interest 
when they have constant spacing. In the case of horizon area, for example, 
one can attempt to relate an equally spaced spectrum (obtained in several 
investigations, e.g., see \cite{pp03} and references therein)
to the intrinsic limitations in  measuring length scales smaller than 
Planck length \cite{padmanabhan87}. 
But the uniform spacing of QNM frequencies is a purely {\it classical} 
result and hence is harder to understand physically. 

It turns out, however,  that one can make {\it some} progress in this 
direction. There are some simple 
derivations, based on the Born approximation, which 
reproduce this structure for the imaginary part 
of the QNM frequencies for a large class of spherically symmetric 
spacetimes like
the Schwarzschild black hole 
\cite{padmanabhan03b, mmv03}. 
The relation obtained, 
valid for large $k$ (i.e., for large $n$),   
has the form
\be
k_n = i n \kappa ~~~ (\mbox{for } n \gg 1)
\e
(In the imaginary part of $k_n$, one obtains 
a factor $n$ instead of $n + (1/2)$ since  the Born approximation 
is expected to be valid only for $n \gg 1$).
The important point to note is that ---  though it is not 
possible to obtain the 
real part of the QNM frequencies using the Born approximation --- 
this simple derivation
does
give the correct level spacing for imaginary values of $k_n$.
Hence the possibility 
of extending the formalism
to spacetimes with more complicated structure 
is worth examining. Following this approach, 
we use the Born approximation to examine the 
level spacing for spacetimes with two horizons.

The simplest spacetime with two horizons is that 
of a black hole in a spacetime with a cosmological 
constant,  
described by the Schwarzschild-DeSitter (SDS) metric. 
The metric is characterized by the presence of a black hole event horizon
and a cosmological horizon. 
In recent times, studying such a spacetime has acquired 
further significance because 
of the cosmological observations suggesting the existence of 
a non-zero positive cosmological constant 
(\cite{pag++99}; for reviews, see \cite{padmanabhan03,ss00}).
While the observations can be explained by a wide class of models 
(see, e.g \cite{pc03}), including
those in which the cosmic equation of state can depend on spatial scale 
\cite{pc02,bjp03}, 
virtually all these models approach the DeSitter (DS) spacetime at late times 
and at large scales. 
Like in the other cases, the QNMs for the SDS spacetime have
been studied both numerically 
(see, for example, \cite{bckl97,bclp99,yf03,cl03,brink03,zhidenko03}) 
and analytically (see, for example, \cite{mm90,mn02,suneeta03}). 
However, since the SDS spacetime is 
characterized by two different surface gravities (corresponding to the 
two horizons), 
the dependence of the level spacing of the 
imaginary part of the QNM frequencies on the surface gravities is not
obvious.
The numerical studies fail to give a ``clean'' result like 
equation (\ref{eq:kn_num}) as they seem to vary depending 
on the relative values of the
two surface gravities. 
A calculation based on the monodromy of the perturbation 
continued to the complex plane \cite{ca03}, similar to what is done for the 
Schwarzschild case, gives a result of the form
$k_n = i \kappa_- \left(n + \f{1}{2}\right) + \mbox{real part}$,
where $\kappa_-$ is the surface gravity of the black hole horizon. 
The real part has a term analogous to $\ln 3$ whose actual form 
depends on the values of the surface gravities. 
The above result is found to be in 
qualitative agreement with numerical results for the near-extremal 
case (i.e., when the surface gravities are very close to each other), 
although the exact numerical coefficients seem to differ 
\cite{yf03}.  
It is not clear 
whether the problem is with the numerics, or the analytical derivation 
misses out on some issues. The above calculation indicates 
that the QNM spectrum is independent of the surface gravity 
of the cosmological horizon $\kappa_+$.
On the other hand the first version of 
\cite{mmv03}, based on Born approximation, claimed
that the spectrum of the imaginary part of the QNMs should behave as 
$k_n = i (n_- \kappa_- + n_+ \kappa_+)$ , where 
$n_{\pm}$ are two integers.  
[This arXiv submission has been revised subsequently, after
the appearance of the current work, removing this claim]. 
Analytical calculation based on approximating 
the scattering potential by some simple form gives 
QNMs proportional to both the surface gravities, depending on the 
time scale one is interested in \cite{suneeta03}. 
Currently, we do not seem to have any 
general consensus on this particular issue!

There are a few more conceptual issues because of which such a study 
(using a prototype metric with two horizons) is important. First is the 
possible  connection between the QNMs and thermodynamics of horizons.
It was pointed out in \cite{padmanabhan03b} that the Born 
approximation result arises from integrals
which are very similar to those which arise in the case of horizon 
thermodynamics (see e.g., equation (23) of \cite{padmanabhan03c}).  
In the case of spacetimes 
with multiple horizons (like SDS), there is no unique temperature 
except when the ratio of surface gravities is a rational number \cite{cp03}. 
If the QNMs
are related to the horizon thermodynamics (in some manner which 
is not yet clearly understood), then
it 
would be interesting to see whether the level spacing in the 
SDS spacetime can give any idea about the temperature of 
spacetimes with multiple horizons. Second, it is well known that 
thermodynamics of gravitating systems
depend crucially on the ensemble which is used 
(see e.g., \cite{padmanabhan90}) which translates into the 
boundary conditions on the horizon (see e.g., \cite{gt03}). 
If QNMs are related to horizon thermodynamics, we will expect some 
similar kind of dependence on the boundary condition for the wave 
modes in case of SDS spacetime.  We shall see that this expectation 
is indeed borne out. Finally, there were also some discussion 
in the literature as to whether the QNMs depend on the region 
beyond the horizons. We will see that we can make some comments 
regarding this issue.

The structure of the paper is as follows. In Sec. II, we briefly 
review the results for the Schwarzschild metric. 
The main problem of interest, the SDS metric, is taken up 
in Sec. III. We derive the explicit form of the scattering amplitude 
using Born approximation and discuss the structure of the 
QNM spectrum. The main conclusions are summarized and 
compared with other results in Sec. IV.

\section{Warm up: QNMs for the Schwarzschild metric}

In this section we review the derivation of the 
QNMs for the Schwarzschild metric using the first Born approximation.
Let us start with a general class of spherically symmetric metrics
of the form
\be
ds^2 = f(r) dt^2 - [f(r)]^{-1} dr^2 - r^2 d\Omega^2
\label{eq:genmet}
\e
with $f(r)$ having the simple zero at $r=r_0$, i.e.,
$f(r) \simeq f'(r_0) (r - r_0)$. 
It was shown in \cite{padmanabhan02c}
that spacetimes described by the above class of metrics
have a fairly straightforward thermodynamic interpretation and -- 
in fact -- Einstein's equations
can be expressed in the form of a 
thermodynamic relation $TdS=dE-PdV$ for such spacetimes, with 
the temperature being determined by the surface gravity of the horizon:
\be
\kappa = \f{1}{2} |f'(r_0)|
\e

Let us consider a massless scalar field $\phi$ satisfying the wave equation 
$\Box \phi = 0$ in this spacetime. 
We look for solutions to the wave equation in the form
\be
\phi= \f{1}{r} F(r) Y_{l m}(\Omega)
e^{ikt};~~ {\rm Re}(k) > 0
\label{eq:sepvar}
\e
Straightforward algebra now leads to a 
``Schrodinger equation'' for $F$ given by:
\be
\left[-\f{d^2}{d r_*^2} + V(r)\right] F(r) = k^2 F(r)
\label{eq:schrodinger}
\e
where the potential is given by
\be
V(r) = f(r) \left[\f{l(l+1)}{r^2} + \f{f'(r)}{r}\right]
\label{eq:pot}
\e
and the tortoise coordinate is defined as
\be
r_* \equiv \int \f{dr}{f(r)}
\e
One can, in principle, 
solve the differential equation (\ref{eq:schrodinger}) given 
a particular set of boundary conditions. However, the equation, in general, 
does not have an exact solution and one has to 
solve it either numerically or by 
using some approximation scheme. 
In this paper, we shall be using 
one such approximation scheme, namely, 
the first Born approximation. 

For a pure Schwarzschild black hole, we have
$f(r) = 1 - 2 M/r$ and the horizon is at $r_0 = 2 M$. The potential 
$V(r)$ in equation 
(\ref{eq:pot}) vanishes at the horizon ($r_* \to -\infty$)
and at spatial infinity ($r_* \to \infty$), which 
means that the wavefunction can be taken to be plane waves in the two  
regions. 
A class of physically acceptable solutions for this system has the 
asymptotic form
\be
F(r) \sim \left\{ \begin{array}{ll}
              e^{i k r_*}
                       & \mbox{(at $r_* \to -\infty$)},\\
              A_{\rm in} e^{i k r_*} + A_{\rm out} e^{-i k r_*}
                       & \mbox{(at $r_* \to \infty$)}.
                \end{array} \right.
\label{eq:sol1}
\e
The ``incident'' wave ($A_{\rm in} e^{i k r_*}$) 
for this class of solutions is propagating 
towards the black hole, and hence the solutions of the above form are
appropriate for studying 
{\it scattering off the black hole}.
The scattering amplitude for the above solution is simply given by
\be
S(k) \propto \f{A_{\rm out}}{A_{\rm in}}
\label{eq:skdef}
\e

Now, the QNMs are defined to be those for which one has a 
purely in-going plane wave at the 
horizon and a
purely outgoing wave at spatial infinity, i.e., which satisfy 
the boundary conditions
\be
F(r) \sim \left\{ \begin{array}{ll}
      e^{i k r_*} & \mbox{(at $r_* \to -\infty$)}, \\
      e^{-i k r_*} & \mbox{(at $r_* \to \infty$)}
      \end{array} \right.
\label{eq:bc1}
\e
It is clear from the equation (\ref{eq:sol1}) that the 
QNMs actually correspond to 
case where $A_{\rm in} = 0$ and are obtained by calculating
the poles of the scattering amplitude
$S(k)$ [see equation (\ref{eq:skdef})]. 
Obtaining the exact form of the scattering amplitude for the 
potential (\ref{eq:pot}) is non-trivial -- hence one has to 
obtain it through some approximation scheme. 
It turns out that one can obtain the explicit form of this
amplitude using the first Born approximation. 

In general, 
the scattering amplitude in the Born approximation is given by the 
Fourier transform
of the potential $V({\bf x})$ with respect to the 
momentum transfer ${\bf q} = {\bf k}_f -{\bf k}_i$:
\be
S({\bf q}) = \int  d{\bf x} ~ V({\bf x}) 
~ e^{-i {\bf q \cdot x}}
\e
In one dimension, ${\bf k}_i$ and ${\bf k}_f$ should be 
parallel or anti-parallel; further we can take their 
magnitudes to be the same for scattering in a fixed potential. 
Then non-trivial momentum transfer occurs only for 
${\bf k}_f = - {\bf k}_i$ so that ${\bf q} = -2 {\bf k}_i$.
From equation (\ref{eq:sol1}), the ``incident'' wave is seen to be
of the form $e^{i k r_*}$, giving the  
the scattering amplitude as 
\bear
S(k) &=& \int_{-\infty}^\infty dr_* V[r(r_*)] e^{2ikr_*}
\nline
&=& \int_{r_0}^{\infty} dr \left[\f{l(l+1)}{r^2} + 
\f{f'(r)}{r}\right] e^{2 i k r_*(r)}
\label{eq:skint}
\ear
where we have omitted irrelevant constant factors. 
(Our original problem was three-dimensional and we are {\it not}
working out the three-dimensional scattering amplitude in, say, $s$-wave 
limit. Rather, we first map the problem to an one-dimensional Schrodinger 
equation and study the scattering amplitude in one dimension). 
This integral picks up significant contribution 
{\it only} near the horizon where 
$r_* \approx (1/2) \kappa^{-1} \ln(r/r_0 - 1)$, 
and it can be shown that the 
approximate form of the scattering amplitude is given by \cite{padmanabhan03b}
\be
S(k) \approx \mbox{constant factors} \times 
\Gamma\left(1 + i \f{k}{\kappa}\right)
\label{eq:born_approx}
\e
It is clear from the above expression that the poles of the 
amplitude is given by the poles of the Gamma function, 
which occur at
\be
k_n  = i n \kappa \qquad ({\rm for}\ n\gg 1)
\label{eq:schpoles}
\e
It also turns out that 
the integral (\ref{eq:skint}) can be solved {\it exactly} for a pure 
Schwarzschild metric \cite{padmanabhan03b}, and the 
imaginary part of the QNM frequencies
are found to be exactly identical to what is obtained above. 
As discussed in Section 1, the Born approximation fails to 
reproduce the real part of the QNM spectrum.

As an aside, we would like to comment on the issue of 
whether QNMs depend on the form of the metric inside the 
horizon and in particular on the singularity of the Schwarzschild 
metric at $r=0$. (These comments do not depend on the Born 
approximation but it is easy to see the result in this limit). 
The answer is essentially ``no'' in the sense that if the Schwarzschild 
metric is modified in a small region around $r=0$, making it 
nonsingular, but leaving the form of the metric unchanged for 
$r\geq 2M$, the QNMs do not change;  but there is subtlety 
in this issue.  It is easy to see that, in the Born approximation, 
we are dealing with scattering problem in the $r_*$ coordinates 
with boundary conditions at $r_*=\pm\infty$. This scattering problem 
only depends on $V[r(r_*)]$ which -- in turn --
depends only on $f(r)$ for $r\geq 2M$. So if we modify the $f(r)$ 
for $r< 2M$, it does not change the Born approximation results. 
What happens when we go beyond the Born approximation and consider 
the real part of QNMs, for example? Here we need to analytically 
continue to the complex values of
$r$ and $r_*$. Now the original definition of the problem, posed 
as a Schrodinger equation in  (\ref{eq:schrodinger}), again cares 
only for the $r_*$ coordinate and is well-defined if boundary conditions 
are specified at $r_*=\pm\infty$. But the relation between 
$r$ and $r_*$ [which again depends only on $f(r)$
at $r\geq 2M$] can be {\it analytically continued for all $r$  
including near $r=0$}. This leads to a unique analytic structure 
in the complex plane  and even 
for $r<2M$ through analytic continuation,  {\it as though} the form
of $f(r)$ is valid all the way to $r=0$! More explicitly, 
this analytical structure is obtained by: (i) defining $f(r)$ for
$r\geq 2M$; (ii) defining $r_*$ using it; (iii) analytically 
continuing this relation to define $r_*$ and $r$ in the complex 
plane and (iv) defining $f(r)$ for all $r$ including near $r=0$
by analytic continuation.  This is independent of the {\it actual} 
form of $f(r)$ for $r<2M$ and previous results like those in 
\cite{mn03,neitzke03} depend only on this structure. This is 
gratifying since we do not know the effects of quantum gravity, 
which could modify the spacetime structure near the singularity.

It might seem that the above formalism can be 
trivially applied to the case
of a spacetime with a cosmological horizon, described by the 
pure DS metric with $f(r) = 1 - H^2 r^2$. The potential 
term for calculating the scattering amplitude, 
given by equation (\ref{eq:pot}), becomes
\be
V(r) = (1 - H^2 r^2) \left[\f{l(l+1)}{r^2} - 2 H^2\right]
\e
which, like in the pure Schwarzschild case, 
vanishes at the horizon $r = H^{-1}$. 
However, in contrast to the Schwarzschild case, 
it does {\it not} vanish at the other boundary, i.e., 
at $r = 0$. It is easy to check that near the origin, 
$V(r) \simeq - 2H^2$ when $l=0$, 
while it blows 
up as $r^{-2}$ for $l > 0$. 
This implies that one cannot take the boundary conditions 
as simple plane waves like in equation (\ref{eq:sol1}). 
In fact, the set of boundary conditions used for studying QNMs in the 
pure DS space is that the wavefunction should be outgoing at the horizon, 
but should vanish at the origin, i.e.,
\be
F(r) \sim \left\{ \begin{array}{ll}
      0 & \mbox{(at $r \to 0$)}, \\
      e^{-i k r_*} & \mbox{(at $r \to H^{-1}$)}
      \end{array} \right.
\label{eq:bcds}
\e 
It turns out that the wave equation can be solved {\it exactly} 
for the DS spacetime \cite{awlq02,acl02}, and for 
scalar fields satisfying the wave equation $\Box \phi = 0$, 
there exist no modes for 
which the above boundary condition is satisfied.  This result is implicit in 
\cite{bclp99}
and we include the details of the calculation in Appendix A for 
completeness.

\section{QNMs for the Schwarzschild-DeSitter metric}

We next consider 
the Schwarzschild-DeSitter (SDS) spacetime, which is described by 
a spherically symmetric metric of the form (\ref{eq:genmet}), with
\be
f(r) = 1 - \f{2M}{r} - H^2 r^2
\e
Let us denote the black hole event horizon and the 
cosmological horizon by $r_-$ and $r_+$ respectively. The corresponding 
surface gravities are denoted as $\kappa_-$ and $\kappa_+$ respectively.
Note that, by definition, both $\kappa_-$ and $\kappa_+$ are 
positive definite. The tortoise coordinate is given by
\bear
r_* 
&=& \f{1}{2 \kappa_-} \ln\left|\f{r}{r_-} - 1\right|
- \f{1}{2 \kappa_+} \ln\left|1 - \f{r}{r_+}\right| 
\nline
&-& \f{1}{2} \left(\f{1}{\kappa_-} -  \f{1}{\kappa_+} \right) 
\ln\left|\f{r}{r_- + r_+} + 1\right|
\nline
\ear
With this definition, the regions $r \le r_-$ and $r \ge r_+$ 
are mapped
to $r_* \le -\infty$ and $r_* \ge \infty$ respectively, 
and we will not require the regions beyond the two horizons. 
(In particular, the form of $f(r)$ inside the Schwarzschild 
horizon and the singularity at $r=0$ are irrelevant in what follows.)
The potential in equation (\ref{eq:pot}) reduces to
\be
V(r) = f(r) \left[\f{l(l+1)}{r^2} + \f{2 M}{r^3} - 2 H^2\right]
\label{eq:pot_sds}
\e
which, because of the $f(r)$ factor, vanishes at both the horizons. 
Thus one can take the boundary conditions to be simple plane waves at the 
two horizons. The usual set of solutions used in a scattering problem is 
identical 
to equation (\ref{eq:sol1}) which, as mentioned earlier, 
is appropriate for studying scattering off the black hole.  
Let us assume that the
boundary conditions which define the QNMs are
still given by 
(\ref{eq:bc1}), which imply that one has purely 
in-going plane waves at the black hole 
horizon ($r = r_-$) and 
purely outgoing waves at the cosmological horizon ($r = r_+$). 
One should realize that this set of boundary conditions implies that the 
waves are propagating ``into'' the horizons at both the 
boundaries and is probably the most reasonable set of  
conditions to be used.

We now apply the first Born approximation, with the ``incident'' 
wave being taken as $e^{i k r_*}$ as before.  
The scattering amplitude is then given by
\be
S(k) = \int_{-\infty}^{\infty} dr_* V(r) e^{2 i k r_*}
\label{eq:sk1}
\e
which can be simplified to
\bear
S(k) &=& 
\int_{r_-}^{r_+} dr \left[\f{l(l+1)}{r^2} + \f{2 M}{r^3} - 2 H^2\right] 
\nline
&\times&
\left(\f{r}{r_-} - 1\right)^{i k/\kappa_-}
\left(1 - \f{r}{r_+}\right)^{-i k/\kappa_+}
\nline
&\times&
\left(1 + \f{r}{r_+ + r_-}\right)^{ik (1/\kappa_+ - 1/\kappa_-)}
\label{eq:skint_sds}
\ear
Like in the pure Schwarzschild case, this integral will pick up 
contribution only near the horizons. However, there are some 
crucial differences which need to be taken care of. Near the 
black hole horizon, we have the usual relation 
$r_* \approx (1/2) \kappa_-^{-1} \ln(r/r_- - 1)$, and the 
contribution is exactly similar to equation (\ref{eq:born_approx}). 
On the other hand, near the cosmological horizon, we have
$r_* \approx - (1/2) \kappa_+^{-1} \ln(1 - r/r_+)$, which 
differs from the other case in the sign of the surface gravity term.
The scattering amplitude will then be a sum of two contributions given by
\bear
S(k) &\approx& \mbox{constant factors} \times 
\Gamma\left(1 + i \f{k}{\kappa_-}\right) \nline
&+& \mbox{constant factors} \times 
\Gamma\left(1 - i \f{k}{\kappa_+}\right)
\label{eq:born_approx_sds}
\ear
In this case, the poles of the amplitude is 
given by the poles of {\it both} the Gamma functions:
\be
k_n = i n \kappa_-, ~~ k_n = -i n \kappa_+ ~~~ (n \gg 1)
\label{eq:sdspoles}
\e
The fact that the amplitude would pick up contributions 
from both the horizons was pointed out earlier in \cite{mmv03}, 
and --- in the first version of \cite{mmv03} that appeared
in the arXiv --- 
it was suggested that the poles would occur at 
$k_n = i (n_- \kappa_- + n_+ \kappa_+)$. 
However, we have shown by the explicit  calculation above 
that this is not correct; summing up 
two contributions to get value of an integral is not the same as adding the
arguments for poles.  
[We note that \cite{mmv03} has since been revised and this particular claim
has been withdrawn]. Further, there is a {\it crucial} sign 
difference in the surface  gravity of the cosmological 
horizon.

This conclusion can be explicitly verified since, fortunately, 
one can evaluate the integral in
(\ref{eq:skint_sds})  {\it exactly}.
Essentially we have to solve integrals of the form
\bear
I_n &=& \int_{r_-}^{r_+} dr ~ r^{-n} \left(\f{r}{r_-} - 1\right)^{i k/\kappa_-}
\left(1 - \f{r}{r_+}\right)^{-i k/\kappa_+}
\nline
&\times&
\left(1 + \f{r}{r_+ + r_-}\right)^{ik (1/\kappa_+ - 1/\kappa_-)}
\ear
with the scattering amplitude being given by the sum
\be
S(k) = 2 M I_3 + l(l+1) I_2 - 2 H^2 I_0
\label{eq:sksum}
\e
The expression for $I_n$ turns out to be \cite{gr94} an integral 
representation of the Appell hypergeometric function $F_1$ 

\bear
I_n &=& (r_+ - r_-)^{1 - ik (1/\kappa_+ - 1/\kappa_-)} 
(r_+ + r_-)^{ik (1/\kappa_- - 1/\kappa_+)}
\nline
&\times&
r_-^{-ik/\kappa_-}
r_+^{ik/\kappa_+}
(r_+ + 2 r_-)^{ik (1/\kappa_+ - 1/\kappa_-)} \nline
&\times&
\f{\Gamma\left(1 + \f{ik}{\kappa_-}\right) 
\Gamma\left(1 - \f{ik}{\kappa_+}\right)}
{\Gamma\left(2 + ik\left[\f{1}{\kappa_-} - \f{1}{\kappa_+}\right]\right)}
\nline
&\times&
r_-^{-n} F_1\left(1 + \f{ik}{\kappa_-},n,
ik\left[\f{1}{\kappa_-} - \f{1}{\kappa_+}\right],
\right.
\nline
&& \left.
2 + ik\left[\f{1}{\kappa_-} - \f{1}{\kappa_+}\right];
-\f{r_+ - r_-}{r_-}, -\f{r_+ - r_-}{r_+ + 2 r_-}\right)
\label{eq:i_n}
\ear 
This expression can be further simplified and written in terms of the 
usual hypergeometric function ${}_2F_1$ for $n = 0,2,3$ -- we give the 
relevant expressions in Appendix B for completeness. The pole 
structure of $I_n$ can, however, be determined from equation (\ref{eq:i_n}) 
itself. The combinations of the form
\be
\f{F_1\left(\alpha,n,\beta,
2 + ik\left[\f{1}{\kappa_-} - \f{1}{\kappa_+}\right];
r_1, r_2\right)}
{\Gamma\left(2 + ik\left[\f{1}{\kappa_-} - \f{1}{\kappa_+}\right]\right)}
\e
which occur in the expression for $I_n$, do {\it not} have any poles. 
(Even though both the denominator and numerator have poles, 
the ratio does not.)
The 
only poles of $I_n$ occur at the poles of the two Gamma functions
$\Gamma\left(1 + \f{ik}{\kappa_-}\right)$ and 
$\Gamma\left(1 - \f{ik}{\kappa_+}\right)$. 
It is then straightforward to see that 
the poles of the scattering amplitude $S(k)$ are given 
by equation (\ref{eq:sdspoles}) -- exactly identical to 
what we obtained by evaluating the integral near the horizons.

Let us now discuss the QNM spectrum as obtained in 
equation (\ref{eq:sdspoles}). Note that 
the QNMs are identified as the positive imaginary 
values of $k$, which implies that the QNMs in this case 
are given by $k_n = i n \kappa_-$; the modes which are dependent 
on $\kappa_+$ correspond to negative imaginary values 
of $k$ and hence do {\it not} represent QNMs. 
[In general, the poles given by negative imaginary values of $k$ correspond to 
bound states of the system. However, since the potential, for $l > 0$,  
(\ref{eq:pot_sds}) is positive everywhere and 
vanishes at the boundaries, it can be shown that there cannot 
exist any bound states for the system \cite{wald79}. Thus, 
the poles given by $k_n = -i n \kappa_+$ are physically irrelevant 
as far as this problem is concerned.]

The above analysis  indicates that the QNMs obtained through the 
first Born approximation  are independent 
of the cosmological horizon.  This conclusion agrees, in the 
large $n$ limit, with 
the imaginary part of the QNM spectrum obtained 
through the monodromy of the perturbation 
continued to the complex plane \cite{ca03}. In view of the fact that
there exists no QNMs for the pure DS spacetime,  ``adding'' 
a black hole near the origin merely introduces the QNMs
corresponding to the black hole. Interpreted in the above manner, this result 
should not be surprising. It is also clear that 
this result gives the two correct limits, i.e., when 
$H \to 0$, the level spacing reduces to that corresponding 
to a Schwarzschild black hole, while for $M \to 0$, we have 
$\kappa_- \to \infty$, and hence there exists no QNMs for the 
pure DS spacetime. 

As an aside, one can also consider 
a different set of boundary conditions, where 
the incident wave is scattered off the cosmological horizon.
It turns out that such conditions will give QNMs proportional 
to $\kappa_+$ [the details of the calculation are given 
in Appendix C]. However, these boundary conditions may not 
be physically relevant and are considered here just 
as a mathematical possibility.

\section{Discussion}

We have used the first Born approximation to obtain the QNM spectrum 
for the SDS spacetime. The approximation gives the 
correct level spacing for the imaginary values of the QNM
frequencies for the Schwarzschild black hole, and the spacing is related to 
the temperature corresponding to the horizon. 
On the other hand, there exist no QNMs for a 
massless  minimally coupled scalar field in a pure DS spacetime. 
It turns out that 
the situation is more complicated in the SDS spacetime and 
depends on the type of scattering one is interested in, 
i.e., on the type of boundary conditions
one imposes.  One can start with the usual set of 
conditions
where an incident wave is propagating towards the black hole and calculate
the scattering amplitude. The poles of the amplitude will then  
represent boundary conditions appropriate for QNMs. It turns out that 
for this case the QNM level spacing depends only on the surface 
gravity of the black hole $\kappa_-$, as expected.
Thus the introduction of a black hole in the 
DS spacetime brings along the appropriate QNMs.

However, there exists another set of boundary 
conditions in which one  
starts with an incident wave propagating towards the cosmological 
horizon. As shown in Appendix C, it is possible to choose the boundary conditions and the
definition of the scattering amplitude such that the QNM level spacing in this 
case depends only on the surface gravity of the cosmological horizon 
$\kappa_+$. We do not believe these boundary conditions 
are physically relevant.

It was found earlier, based on the monodromy of the perturbation 
continued to the complex plane \cite{ca03} that the 
imaginary part of the QNM frequencies have an equally spaced structure, 
with the spacing dependent only on $\kappa_-$.
It is not clear whether there exists any extensions of the 
above procedure for obtaining the other set of QNMs which are dependent
on $\kappa_+$. Analytical 
calculations, based on approximating the potential 
by a Poschl-Teller form \cite{suneeta03}, gave 
a QNM spectrum which depends on {\it both} the surface gravities 
$\kappa_-$ and $\kappa_+$ 
depending on the time-scale one is interested in. 
Since we are studying a time-independent 
situation, it is difficult to comment on time-dependence of the
QNM spectrum -- however, our analysis indicates that if one starts 
with an incident wave packet which is composed of monochromatic
waves propagating in both 
directions (i.e., terms of the form $e^{i k r_*}$ and $e^{-i k r_*}$), 
then one might obtain a QNM spectrum which 
depends on both the surface gravities. There is one more crucial difference 
between our results and those obtained by the 
Poschl-Teller potential \cite{suneeta03}-- the level 
spacing in the later case is $2 \kappa_{\pm}$ rather than 
$\kappa_{\pm}$. Such a difference was noted in the 
case of QNMs obtained by Born approximation for the 
Schwarzschild black hole  
\cite{padmanabhan03b} while comparing with results obtained by approximating 
the potential \cite{solodukhin03}; the difference is probably 
related to the incorrect use of $q=k_i$ rather than 
$q = 2 k_i$ for momentum transfer
in the Born approximation.

Most of the numerical computations regarding the SDS spacetime 
concentrate on the near extremal case (where $\kappa_+ 
\approx \kappa_-$), and it is found that the imaginary 
part of the QNM frequencies have an equally spaced structure with 
the spacing given by either of the surface gravities 
(which are anyway equal to the lowest order)
\cite{cl03,brink03,yf03}. 
This means that, to the lowest order, we do not find any 
disagreement between our results and numerical computations.
In other numerical 
computations, where the values of the 
two surface gravities are taken to be widely different 
\cite{bckl97,bclp99}, one obtains 
two sets of QNM spectra proportional to 
the two surface gravities, each valid at different time-scales. 
At this stage, we are unable to make any direct comparison with 
such results since the issue of time-scales is 
difficult to settle in our approach.

In future studies for the SDS spacetime, 
it would be interesting to calculate the scattering amplitude
using some more rigorous technique (like, say, what 
is done for the pure Schwarzschild case \cite{neitzke03}) and 
see how the real parts of the QNM frequencies depend on the 
different surface gravities and on the type of scattering.

\appendix

\section{QNMs for the pure DeSitter metric}
\vspace{-0.25cm}

In this appendix, we give the details of the calculations for 
calculating the QNMs for the DS spacetime. Although most of 
the mathematical apparatus already 
exists in literature \cite{lp78,polarski89,polarski89b,bclp99,awlq02,acl02}, 
we include the details 
for completeness and for emphasizing the conclusion.

The radial wave equation for the DS metric 
[see equations (\ref{eq:schrodinger}) and (\ref{eq:pot})]
\bear
&&\left[-\left\{(1 - H^2 r^2) \f{d}{d r}\right\}^2 \right.
\nline
&& +\left.
(1 - H^2 r^2) \left\{\f{l(l+1)}{r^2} - 2 H^2\right\}\right] F(r) = k^2 F(r)
\nline
\ear
can be reduced 
to the  hypergeometric form by introducing a new variable $z = r^2 H^2$. 
The solution, {\it which is regular 
at the origin}, can be written as
\bear
F(r) &=& r^{l+1} (1 - H^2 r^2)^{i k/2 H} 
\nline
&\times&
{}_2F_1\left(\f{l}{2} + \f{i k}{2 H}, \f{l}{2} + \f{3}{2} + \f{i k}{2 H}, 
l + \f{3}{2}; H^2 r^2\right)
\nline
\ear
where the normalization is arbitrary.  The behaviour of this solution 
near the horizon $r = H^{-1}$ is given by
\bear
F(r) &\propto& 
\Gamma\left(l + \f{3}{2}\right) 
\nline
&\times&
\left[(1 - H^2 r^2)^{-i k/2 H} \f{\Gamma\left(\f{i k}{H}\right)}
{\Gamma\left(\f{l}{2} + \f{i k}{2 H}\right) 
\Gamma\left(\f{l}{2} + \f{3}{2} + \f{i k}{2 H}\right)}
\right. 
\nline
&+& \left.(1 - H^2 r^2)^{i k/2 H} \f{\Gamma\left(-\f{i k}{H}\right)}
{\Gamma\left(\f{l}{2} - \f{i k}{2 H}\right) 
\Gamma\left(\f{l}{2} + \f{3}{2} - \f{i k}{2 H}\right)}
\right]
\nline
\ear
According to the boundary conditions (\ref{eq:bcds}), 
the solution should be purely 
outgoing near the horizon, i.e., 
$F(r) \sim (1 - H^2 r^2)^{i k/2 H}$. [This follows from the fact that
$r_* = H^{-1} \tanh^{-1} (Hr)$ for the DS metric which, near the horizon, 
gives $1 - H^2 r^2 = \sech^2(H r_*) \sim e^{-2 H r_*}$.] This 
implies that the QNMs are given by the poles of the expression
\be
\f{\Gamma\left(\f{l}{2} + \f{i k}{2 H}\right) 
\Gamma\left(\f{l}{2} + \f{3}{2} + \f{i k}{2 H}\right)}
{\Gamma\left(\f{i k}{H}\right)}
\e
The numerator has two sets of poles at 
$k_{n,l} = i H (2n + l)$ and $k_{n,l} = i H (2n + l + 3)$ for 
$n = 0,1,2,...$. However, each of these poles is canceled 
by a similar pole of the Gamma function in the denominator 
\cite{bclp99}. Hence, there exist no QNMs for the 
pure DS spacetime which obey the boundary condition (\ref{eq:bcds}).

Note that this conclusion is only true for a 
{\it massless, minimally coupled} scalar field 
with wave equation $\Box \phi = 0$.  There do exist
well-defined QNMs for a massive scalar field, or for a
scalar field coupled to the Ricci scalar.

\section{Simplified expressions for the scattering amplitude}
\vspace{-0.25cm}

In this appendix, we shall write the scattering amplitude given by 
equations (\ref{eq:sksum}) and (\ref{eq:i_n}) in terms of the 
more familiar hypergeometric functions.
For notational convenience, let us define
\be
a_k \equiv \f{k}{\kappa_-}, b_k \equiv \f{k}{\kappa_+}, 
c_k \equiv a_k - b_k = k \left(\f{1}{\kappa_-} -  \f{1}{\kappa_+}\right)
\e
Also define
\be
A_k = (r_+ - r_-)^{1 + i c_k} 
(r_+ + r_-)^{i c_k}
r_-^{-i a_k}
r_+^{i b_k}
(r_+ + 2 r_-)^{-i c_k} 
\e
Then we have a much simpler expression:
\bear
I_n &=& A_k
\f{\Gamma\left(1 + i a_k\right) 
\Gamma\left(1 - i b_k\right)}
{\Gamma\left(2 + i c_k\right)}
\nline
&\times&
r_-^{-n} 
F_1\left(1 + i a_k, n,
i c_k,
2 + i c_k;
\right.
\nline
&& \left.
-\f{r_+ - r_-}{r_-}, -\f{r_+ - r_-}{r_+ + 2 r_-}\right)
\ear 
For calculating the scattering amplitude, 
we are only interested in the three quantities $I_0, I_2, I_3$. 
For $n=0$, use the relation
\be
F_1(a,0,b,c,x,y) = {}_2F_1(a,b,c;y)
\e
to obtain
\bear
I_0 &=& A_k
\Gamma\left(1 + i a_k\right) 
\Gamma\left(1 - i b_k\right)
\nline
&\times&
\f{{}_2F_1\left(1 + i a_k,
i c_k,
2 + i c_k;
-\f{r_+ - r_-}{r_+ + 2 r_-}\right)}
{\Gamma\left(2 + i c_k\right)}
\label{i0}
\ear
Similarly, for $n=2$, use the relation
\be
F_1(a,c-b,b,c,x,y) = (1-x)^{-a} {}_2F_1\left(a,b,c;\f{y-x}{1-x}\right)
\e
to write
\bear
&&\!\!\!\!\!\!F_1\left(1 + i a_k, 2,
i c_k,
2 + i c_k;
-\f{r_+ - r_-}{r_-}, -\f{r_+ - r_-}{r_+ + 2 r_-}\right)
= 
\nline
&&\!\!\!\!\!\!
\left(\f{r_+}{r_-}\right)^{-1-ia_k} 
{}_2F_1\left(1 + i a_k,i c_k,2 + i c_k;\f{r_+^2 - r_-^2}{r_+(r_+ + 2 r_-)}\right)
\nline
\ear
and hence
\bear
I_2 &=& A_k
\Gamma\left(1 + i a_k\right) 
\Gamma\left(1 - i b_k\right)
r_-^{-2} \left(\f{r_+}{r_-}\right)^{-1-ia_k}
\nline
&\times&
\f{ 
{}_2F_1\left(1 + i a_k,i c_k,2 + i c_k;\f{r_+^2 - r_-^2}{r_+(r_+ + 2 r_-)}\right)}
{\Gamma\left(2 + i c_k\right)}
\nline
\label{i2}
\ear 
Finally, use
\bear
&&\!\!\!\!\!\!F_1(a,c-b+1,b,c,x,y) = (1-x)^{-a} 
\left[{}_2F_1\left(a,b,c;\f{y-x}{1-x}\right)
\right.
\nline
&&\!\!\!\!\!\!+
\left.
\f{a}{c} \f{x}{1-x} {}_2F_1\left(a+1,b,c+1;\f{y-x}{1-x}\right)\right]
\nline
\ear
to obtain
\bear
&&\!\!\!\!\!\!F_1\left(1 + i a_k, 3,
i c_k,
2 + i c_k;
-\f{r_+ - r_-}{r_-}, -\f{r_+ - r_-}{r_+ + 2 r_-}\right)
= 
\nline
&&\!\!\!\!\!\!
\left(\f{r_+}{r_-}\right)^{-1-ia_k} \left[ 
{}_2F_1\left(1 + i a_k,i c_k,2 + i c_k;\f{r_+^2 - r_-^2}{r_+(r_+ + 2 r_-)}\right)
\right.
\nline
&&\!\!\!\!\!\! - (1 + i a_k) \f{r_+ - r_-}{r_+} 
\nline
&& \left. \times
\f{{}_2F_1\left(1 + i a_k,i c_k,3 + i c_k;\f{r_+^2 - r_-^2}{r_+(r_+ + 2 r_-)}\right)}
{2 + i c_k}\right]\nline
\ear
and hence
\bear
I_3 &=& A_k
\Gamma\left(1 + i a_k\right) 
\Gamma\left(1 - i b_k\right)
r_-^{-3} \left(\f{r_+}{r_-}\right)^{-1-ia_k}
\nline
&\times&
\left[
\f{ 
{}_2F_1\left(1 + i a_k,i c_k,2 + i c_k;\f{r_+^2 - r_-^2}{r_+(r_+ + 2 r_-)}\right)}
{\Gamma\left(2 + i c_k\right)}
\right.
\nline
&-& 
(1 + i a_k) \f{r_+ - r_-}{r_+} 
\nline
&&
\left.
\times
\f{{}_2F_1\left(1 + i a_k,i c_k,3 + i c_k;\f{r_+^2 - r_-^2}{r_+(r_+ + 2 r_-)}\right)}
{\Gamma(3 + i c_k)}
\right]\nline
\label{i3}
\ear 

To determine the pole structures of the 
three quantities $I_0, I_2, I_3$, note that all of them 
contain 
combinations of the form 
\be
\f{{}_2F_1\left(1 + i a_k,i c_k,n + i c_k;
r_1\right)}
{\Gamma\left(n + i c_k\right)}
\e
One might notice that both the 
quantities ${}_2F_1\left(1 + i a_k,i c_k,n + i c_k;r_1\right)$ and 
$\Gamma\left(n + i c_k\right)$ have poles at negative integral 
values of $n + i c_k$ -- however, their ratio turns out to 
be regular everywhere \cite{gr94}. This implies that the 
poles of $I_0, I_2, I_3$ occur only at the poles of the Gamma functions
$\Gamma(1 + i a_k)$ and $\Gamma(1 - i b_k)$, 
which is what we have been using in the main 
text.

\section{QNMs for a different set of boundary conditions}
\vspace{-0.25cm}

One should note that the solutions (\ref{eq:sol1}) used 
are appropriate for describing scattering 
off the black hole. As a mathematical possibility, 
it might be interesting to see what happens if one uses a different 
boundary condition in the form of the 
solutions which represent the {\it scattering off the 
cosmological horizon}, i.e., 
\be
F(r) \sim \left\{ \begin{array}{ll}
        \bar{A}_{\rm in} e^{-i k r_*} + \bar{A}_{\rm out} e^{i k r_*} 
             & \mbox {(at $r_* \to -\infty$)}, \\
        e^{-i k r_*} & \mbox {(at $r_* \to \infty$)}
        \end{array} \right.
\label{eq:sol2}
\e
Note that in this case, the ``incident'' wave is propagating towards 
the cosmological horizon.
For the pure black hole case, these solutions are 
irrelevant as they correspond to the 
physically unacceptable case 
where the incident waves propagate towards spatial infinity. 
These solutions are not relevant for the pure DS case too; 
this is because the potential 
is non-zero at $r=0$ and hence it is never possible 
to have an in-going plane wave at this region. 
However, in the case of the SDS spacetime,   
one might consider the above case for studying scattering off
the cosmological horizon at $r_* \to \infty$. 
Clearly, the scattering amplitude in this case will be 
exactly like equation (\ref{eq:skdef}), i.e., 
$S(k) \propto \bar{A}_{\rm out}/\bar{A}_{\rm in}$, and the 
QNMs, still defined by the boundary conditions (\ref{eq:bc1}), 
are given by the poles of the scattering 
amplitude. 
Setting $\bar{A}_{\rm in}=0$ in (\ref{eq:sol2}) leads to the same boundary
condition (\ref{eq:bc1}) as obtained by setting 
$A_{\rm in}=0$ in (\ref{eq:sol1}). 
In that case, we will obtain the same poles for the 
scattering matrix as before.
If, on the other hand, we also change the sign of ${\bf k}_i$ 
in the definition of scattering amplitude when we consider 
the ``incident" wave travelling in the opposite direction, 
then poles flip sign. In this case,
the scattering 
amplitude in the first Born approximation is given by
\be
S(k) = \int_{-\infty}^{\infty} dr_* V(r) e^{- 2 i k r_*}
\label{eq:sk2}
\e
Note that because of the different form of the ``incident'' wave, 
the sign of $k$ in the above equation is different from the 
corresponding equation (\ref{eq:sk1}) in the previous case. The analysis 
of Sec. III goes through identically, except that 
{\it there is a crucial difference in the sign of $k$}.  
The explicit form of $S(k)$ can be calculated
exactly as in Sec. III, and because of the change 
in the sign of $k$, one would find that the poles occur at
\be
k_n = - i n \kappa_-, ~~ k_n = i n \kappa_+ ~~~ (n \gg 1)
\e
This shows that the QNMs, given by positive 
imaginary $k$, are now determined by the 
surface gravity of the cosmological horizon, as expected.
While this is certainly a mathematical possibility, 
the boundary conditions in (\ref{eq:sol2})
seems artificial from the physical point of view.

\bibliography{astropap}

\end{document}